\documentclass[twocolumn,pra,showpacs,floatfix,amsmath,amsfonts]{revtex4-1}
\usepackage[latin9]{inputenc}
\usepackage{color}
\usepackage{amsmath}
\usepackage{amssymb}
\usepackage{graphicx}
\usepackage{esint}

\makeatletter
\@ifundefined{textcolor}{}
{%
 \definecolor{BLACK}{gray}{0}
 \definecolor{WHITE}{gray}{1}
 \definecolor{RED}{rgb}{1,0,0}
 \definecolor{GREEN}{rgb}{0,1,0}
 \definecolor{BLUE}{rgb}{0,0,1}
 \definecolor{CYAN}{cmyk}{1,0,0,0}
 \definecolor{MAGENTA}{cmyk}{0,1,0,0}
 \definecolor{YELLOW}{cmyk}{0,0,1,0}
}

\usepackage{color}
\usepackage{graphics}
\usepackage{epsfig}
\newcommand{\be}{\begin{equation}}
\newcommand{\ee}{\end{equation}}
\newcommand{\bea}{\begin{eqnarray}}
\newcommand{\eea}{\end{eqnarray}}
\newcommand{\bes}{\begin{subequations}}
\newcommand{\ees}{\end{subequations}}
\newcommand{\PT}{\mathcal{PT}}

\newcommand{\tV}{\tilde{V}}

\newcommand{\tc}{\tilde{c}}

\newcommand{\p}{{\cal P}}
\newcommand{\T}{{\cal T}}
\newcommand{\hal}{\hat{\alpha}}

\newcommand{\bc}{\mathbf{c}}

\newcommand{\bPsi}{\boldsymbol{\Psi}}

\newcommand{\cV}{\mathcal{V}}
\newcommand{\cH}{\mathcal{H}}

\makeatother

\begin{document}

\title{Dynamical suppression of tunneling and spin switching of a spin-orbit-coupled atom in a double-well trap
}

\author{Y. V. Kartashov$^{1,2}$, V. V. Konotop$^{3}$,   and V. A. Vysloukh$^4$}

\affiliation{ $^1$ICFO-Institut de Ciencies Fotoniques, The Barcelona Institute of Science and Technology, 08860 Castelldefels (Barcelona), Spain 
	\\
	$^2$Institute of Spectroscopy, Russian Academy of Sciences, Troitsk, Moscow, 108840, Russia
	\\
 $^{3}$Centro de F\'{i}sica Te\'orica e Computacional  and Departamento de F\'{i}sica, Faculdade de Ci\^encias, Universidade de Lisboa, Campo Grande, Ed. C8, Lisboa 1749-016, Portugal
\\
$^4$Departamento de Fisica y Matematicas, Universidad de las Americas Puebla, 72820 Puebla, Mexico
}

\date{\today}
\begin{abstract}
We predict wide-band suppression of tunneling of spin-orbit-coupled atoms (or noninteracting Bose-Einstein condensate) in a double-well potential with periodically varying depths of the potential wells. The suppression of tunneling is possible for a single state and for superposition of two states, i.e. for a qbit. By varying spin-orbit coupling  one can drastically increase the range of modulation frequencies in which an atom remains localized in one of the potential wells, the effect connected with crossing of energy levels. This range of frequencies is limited because temporal modulation may also excite resonant transitions between lower and upper states in different wells. The resonant transitions enhance tunneling and are accompanied by pseudo-spin switching. Since the frequencies of the resonant transitions are independent of potential modulation depth, in contrast to frequencies at which suppression of tunneling occurs, by varying this depth one can dynamically control both spatial localization and pseudo-spin of the final state.
\end{abstract}

\pacs{...}

\maketitle

\section{Introduction}

Tunneling is a fundamental quantum phenomenon determining principles of operation of the majority of devices utilizing electrons, nucleus, atoms, or molecules. Therefore control of tunneling is an important task that continuously attracts considerable attention in diverse areas of physics. More than two decades ago it was discovered~\cite{Gross} that periodic time modulations of the parameters of the potential, in which a quantum particle is trapped, is an efficient tool for control of tunneling that allows its suppression for certain values of modulation frequencies. Nowadays, the control of tunneling in scalar systems driven by periodic forces is a well studied topic of quantum mechanics~\cite{DrivenTun}. Suppression of tunneling was experimentally observed in optical settings based on waveguides modulated in the direction of light propagation~\cite{Optics1} (see~\cite{Optics3} for review).

In recent years, technological advances in manipulation of particles with spin and development of spintronics, brought to light novel problems and perspectives related to spin-dependent tunneling~\cite{spintronics}.  Just like tunneling of different quantum particles, say electrons, atoms, photons, etc. can be considered in the unified mathematical framework, the tunneling of spinors, for which either spin or quasi-spin play the role of additional internal degrees of freedom, allow for general description irrespectively of their physical nature.  Due to this additional degree of freedom, tunneling may acquire new features. Such systems, realized as Bose-Einstein condensates (BECs) of mixtures of atomic hyperfine states, were already studied experimentally~\cite{Willams,Oberthaler}. Since double-well potential is the most common model for studying tunneling~\cite{LL}, spinor BECs in such potentials received considerable attention~\cite{spinor}. 

The idea of simulation of spin-bearing particles by neutral multilevel atoms was further developed after discovery and experimental realization of spin-orbit-coupled (SOC) cold atoms and BECs~\cite{Spielman}. 
%~\cite{Spielman,reviews_SO}.
 Recently, SOC-BECs in double-well potentials were addressed by several authors with particular emphasis on the dynamics of Josephson oscillations~\cite{SOC-double-well}, self-trapping~\cite{SOC-self-trap}, and on the effects stemming from energy level crossing~\cite{double_flat}. However, possibility of dynamical suppression of tunneling in SOC-systems have never been studied. 

In this work we report on dynamical suppression of tunneling of a SOC atom (or noninteracting BEC) in a double-well potential, whose wells exhibit out-of-phase periodic modulations. SOC has profound effect on tunneling dynamics and may lead to considerable increase of modulation frequency intervals, at which an atom remains trapped in one well. This effect is a counterpart of the dynamic localization~\cite{Dunlap} reported for SOC atoms in~\cite{KKZT}, and also based on the existence of  flat bands in lattices~\cite{YoChu,SOC-Bloch}. Even though SOC-induced energy crossing can occur only for one out of two pairs of energy levels, that distinguishes a realistic system from early reported simplified models~\cite{double_flat}, periodic modulation allows to achieve crossing of quasi-energies (leading to localization) also for other pair of levels. This interplay between crossing of the levels of a stationary potential and dynamical crossing of Floquet exponents, enables localization of superposition of states, which can be viewed as a qbit, in a single potential well. Certain modulation frequencies induce resonant transitions between lower and higher localized states in different potential wells. These resonances limit the intervals of frequencies at which suppression of tunneling occurs. Since by changing the amplitude of modulation the point of crossing of quasi-energies can be shifted at will, one can observe the phenomenon of spin flipping occurring when quasi-energy crossing occurs exactly at the frequency of resonant transition between lower and upper levels. In this case spin flipping accompanies tunneling of an atom between the potential wells.

The paper is organized as follows. In Sec.~\ref{sec:model} we formulate the model and derive four mode approximation. The dynamical suppression of tunneling and associated effects are described in Sec.~\ref{sec:tunneling}. In Sec.~\ref{sec:nonlinearity} the effect of weak nonlinearity of the suppression of tunneling is reported. The outcomes are summarized in the concluding section. 

\section{The continuous model and its four mode approximation}
\label{sec:model}

We consider a two-level atom described by the spinor $\boldsymbol{\Psi}=(\Psi_{1},\Psi_{2})^{T}$, which is trapped in a one-dimensional double-well potential $V(x)$. The trap is produced by a sum of two identical potentials $V_0(x)=V_0(-x)$ located at $x=\pm d$, i.e., $V(x)=V_-(x)+V_+(x)$ where $V_\pm (x)=V_0(x\mp d/2)$ and $V_0(x)\to0$ at $|x|\to\infty$. The tunneling is controlled by small-amplitude, periodic, out-of-phase modulation of depths of the potential wells, described by $f\sin(\omega t)\tV(x)$, where $\tV(x)=-V_-(x)+V_+(x)$, $\omega$  is the modulation frequency, and $f\ll1$ is the amplitude of modulation. We denote the strength of SOC as $\gamma$  and choose the units in which $\hbar=m=1$. In this  case, the Hamiltonian of our continuous model takes the form $H=H_0+f\sin(\omega t)\tV(x)$, where 
\begin{eqnarray}
H_0=\frac{p^2}{2}-\gamma\sigma_{z}p+\Omega\sigma_{x}+V(x)
\end{eqnarray}
  is the unperturbed Hamiltonian of a SOC atom in a double-well potential, $2\Omega$ is the Zeeman spitting, and $\sigma_{x,y,z}$ are the Pauli matrices.

 \begin{figure}[h]
\includegraphics[width=\columnwidth]{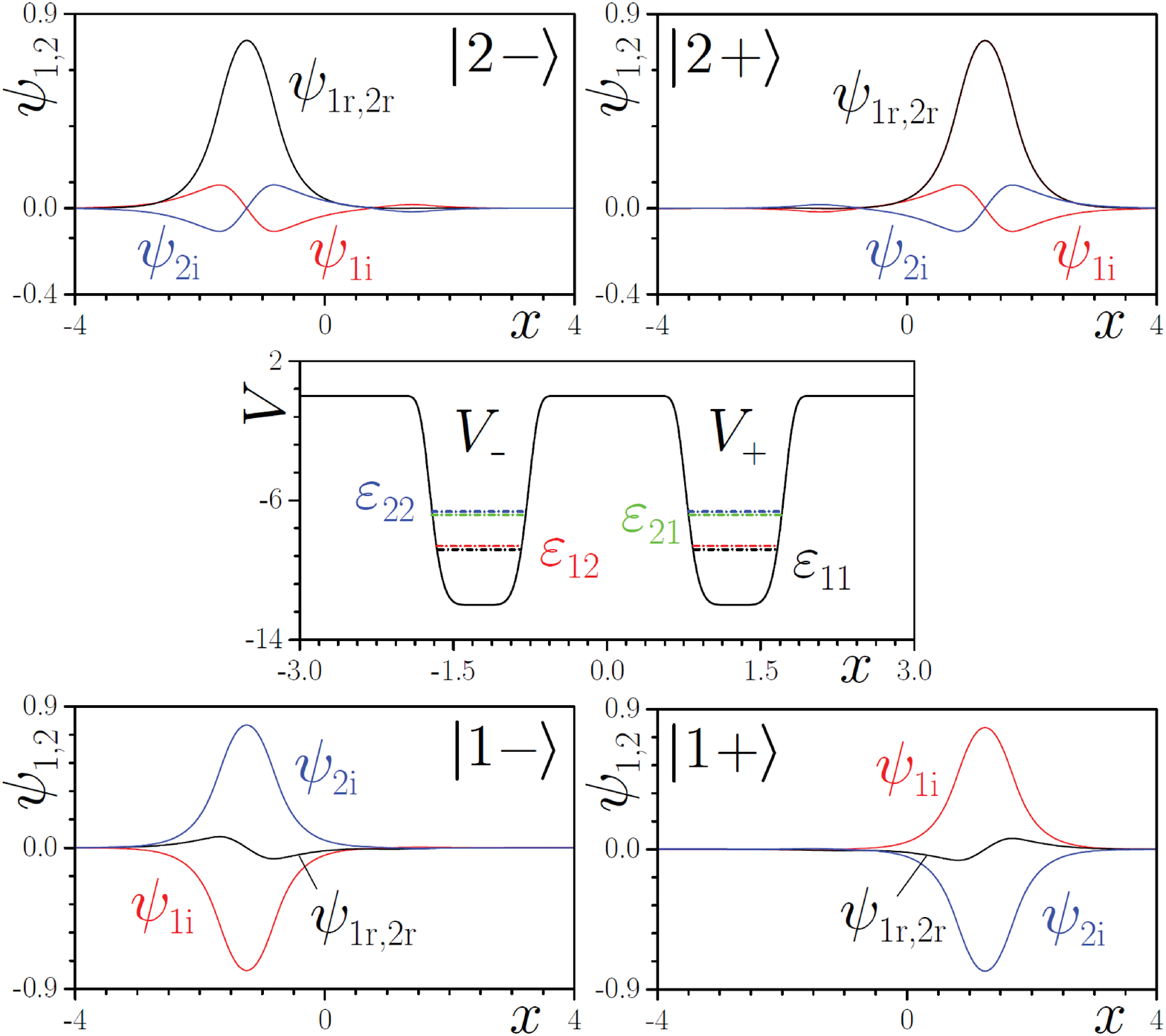}
\caption{{The shape of the double-well potential and localized} modes (indicated in top and bottom panels) at $\gamma=0.8$. {The energy levels corresponding to different modes are indicated by dashed lines in the panel with potential}. Subscripts "r" and "i" stand for the real and imaginary parts of the respective functions.}
\label{fig:one}
\end{figure}

It is assumed that each well $V_0(x)$, {being considered separately in the presence of SOC,} possesses \textit{two} discrete eigenstates, whose splitting {due to coupling with neighbouring well} yields four states $|ij\rangle$ ($i,j=1,2$) of the entire stationary double-well potential: $H_0|ij\rangle=\varepsilon_{ij} |ij\rangle$. In each state the first index $i=1,2$ refers to the lower/upper pair of levels, while the second index  $j=1,2$ refers to the lower/upper levels in each pair. According to this nomenclature $\varepsilon_{11}<\varepsilon_{12}<\varepsilon_{21}<\varepsilon_{22}<0$.
 
The unperturbed Hamiltoninan $H_0$ obeys three symmetries:  $\hat{\alpha}_1=\PT$, $\hat{\alpha}_2=\sigma_x\T$, and $\hat{\alpha}_3=\sigma_x\p$ with $\p$ and $\T$ being space and time reversal operators. The operators $\hal_{1,2,3}$ together with the identity operator constitute a Klein four-group characterized by the properties $\hat{\alpha}_n\hat{\alpha}_n=1$ and $\hat{\alpha}_m\hat{\alpha}_n=\hat{\alpha}_k$. Thus, the eigenstates $|ij\rangle$ can be chosen to obey all $\hal$'s symmetries, i.e., $\hal_{n}|ij\rangle=|ij\rangle$ for $n=1,2,3$. Namely these states are considered below.  Using the symmetry properties one can prove that the average $x-$component of the spin is the only nonzero component for all eigenstates, i.e. $\langle ij|\sigma_{x}|ij\rangle\neq 0$  and $\langle ij|\sigma_{y,z}|ij\rangle=0$.
 
Next we introduce a new orthonormal basis $|i\pm\rangle=\frac{1}{\sqrt{2}}\left(|i1\rangle\pm |i2\rangle\right)$, whose functions are localized in the left ("$-$") or in the right ("$+$") potential wells (see Fig.~\ref{fig:one}). The modes in Fig.~\ref{fig:one} are calculated for the potential with $V_0(x)=-U\exp(-x^6/a^6)$, depth $U=12$ and width $a=1/2$. The separation between wells in $V(x)$ is given by $d=2.5$.  The frequency $\Omega$ was set to $1$, its {specific value, as long as it remains of the order of $1$,} does not  affect qualitatively the phenomena described below. The result of the action of the symmetry operators introduced above on these states can be written as  
\begin{equation}
\hat{\alpha}_1|{i}\pm\rangle =|{i}\mp\rangle, \,\,\hat{\alpha}_2|{i}\pm\rangle   =|{i}\pm\rangle,\,\,   \hat{\alpha}_3|{i}\pm\rangle  =|{i}\mp\rangle
\end{equation}
{\color{black}(note that kets $|{i}\pm\rangle$ are not eigenstates of $H_0$ anymore)}. The imposed symmetries imply that the $x-$projections of the mean spins of the lower and upper modes, $S_{xi}=\langle i\pm|\sigma_x|i\pm \rangle/2 $, are equal for both wells but have opposite directions: $S_{x1}\approx-0.4878$, $S_{x2}\approx 0.4693$ for $\gamma=0.8$ and $S_{x1}\approx-0.4584$, $S_{x2}\approx 0.4059$  for $\gamma=1.5$. Thus, each such mode represents nearly pure state.

By neglecting the transitions to the continuous spectrum one can describe the evolution of any state by the spinor 
\begin{eqnarray}
\bPsi=e^{-iE_0t}\sum_{i=1}^2\left( c_{i-}(t)|i-\rangle+ c_{i+}(t)|i+\rangle\right)
\end{eqnarray}
 where $E_0=\frac{1}{4}\sum_{ij}\varepsilon_{ij}$  and evolution of the column-vector $\bc= (c_{1-},c_{1+},c_{2-},c_{2+})^T$ (here "T" stands for transpose) is described by the system 
   \begin{eqnarray}
   \label{eq:matrix_main}
   i\frac{d\bc}{dt}=(\cH_0+\cH_\delta)\bc+f\sin(\omega t)\cV\bc.  
    \end{eqnarray}
 Here $\cH_0=\Delta\,$diag$(-1,-1,1,1)$, $$
 \Delta=\frac{1}{4}(\varepsilon_{22}+\varepsilon_{21}-\varepsilon_{12}-\varepsilon_{11})
 $$ 
 describes dephasing of the lower and upper modes, $\cH_\delta=$diag$(\delta_1\sigma_x,\delta_2\sigma_x)$ with $\delta_i=(\varepsilon_{i2}-\varepsilon_{i1})/2$ is the $4\times4$ block matrix describing Josephson tunneling of the states between the wells due to nonzero energy difference. The matrix $\cV$ accounting for modulation has the form  
\begin{eqnarray}
\cV=\left(\begin{array}{cc}
v_1\sigma_{z} & u\sigma_z+iw\sigma_y
\\ 
 u\sigma_z-iw\sigma_y & v_2\sigma_{z} 
\end{array}\right)
\end{eqnarray}
where $v_i=\langle i-|(V_+-V_-)|i-\rangle$ is the energy deviation of the $i-$th state from the mean value $\pm\Delta$, $u=\langle 1-|(V_+-V_-)|2-\rangle$ describes the probability of transitions between states residing in the same potential well, while $w=\langle 1-|(V_--V_+)|2+\rangle$ describes transitions between lower state in one well and upper state in other well. The latter are termed here Josephson-Rabi transitions, since they simultaneously involve transitions between lower and upper energy levels and tunneling between potential wells. The coefficients $v_{1,2}$, $u$, and $w$ are all real. The orthonormality of the modes $|i\pm\rangle$ implies relatively small probability of the Rabi transitions $|1\pm\rangle \leftrightarrow |2\pm\rangle$ in comparison with Josephson-Rabi ones $|1\pm\rangle \leftrightarrow |2\mp\rangle$, i.e. $|u|\ll|w|$. In particular, in all cases considered below $|u/w|\sim 10^{-3}$.

\section{Dynamical suppression of tunneling}
\label{sec:tunneling}

Dynamical suppression of tunneling is achieved when modulation frequency $\omega$ and amplitude $f$ are such that \textit{single} initial state, say $|\psi\rangle=|i-\rangle$, remains single over large time interval. This implies that no coupling with other modes occurs.  We refer to this case as to \textit{partial} suppression of tunneling. \textit{Full} suppression of tunneling takes place when for an input in a form of superposition of modes from \textit{both} upper and lower levels, $|\psi\rangle=c_{1-}|1-\rangle+c_{2-}|2-\rangle$ only the amplitudes  $c_{i-}(t)$ remain nonzero. Thus, the dynamical suppression of tunneling within the framework of four-mode model (\ref{eq:matrix_main}) is expected to be fully determined by its Floquet exponents $\lambda$ (quasi-energies). Modulation frequencies, at which Floquet exponents cross, correspond to suppression of tunneling of the respective states. In terms of the full Schr\"odinger equation $i\partial\bPsi /\partial t=H\bPsi$, with the time-dependent Hamiltonian $H$, the dynamical suppression of tunneling can be quantified by the time-averaged probability 
\begin{eqnarray}
P_<(t)=\frac{1}{t} \int_0^tdt'\int_{-\infty}^{0}dx\bPsi^\dagger(x,t')\bPsi(x,t')
\end{eqnarray} 
of finding an atom at $x<0$, where averaging is performed over sufficiently long time interval. {\color{black}  Specifically we have chosen  $t=1000$ in all numerical simulations. For rubidium atoms in the trap whose single-well width is $4\,\mu$m (this corresponds to $a=1/2$ in the super-Gaussian model used in our simulations), the dimensionless unit of time corresponds to approximately 1 ms in the physical units. Thus averaging is performed over approximately $1\,$s. Respectively, the frequencies shown in all figures below are measured in the units of $375.5\,$s$^{-1}$.} 

If the initial state is localized in the left potential well, the tunneling is suppressed and spinor remains localized in the left well for almost all moments of time $t'<t$ for $P_<(t)$ close to $1$. If the system undergo periodic Josephson or Josephson-Rabi oscillations, then $P_<(t)$ is close to $1/2$.

\begin{figure}[h]
 \includegraphics[width=\columnwidth]{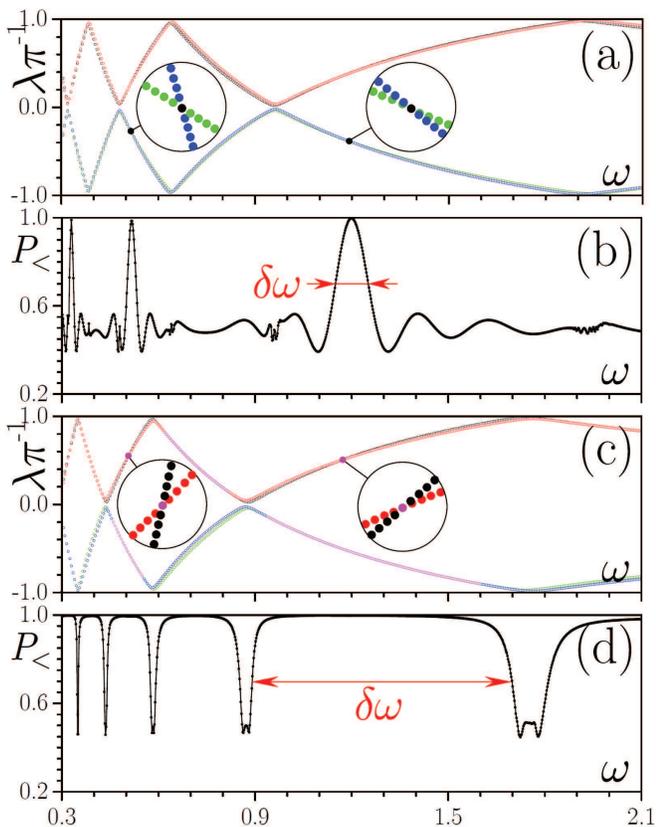}
 \caption{ Floquet exponents $\lambda$ (a),(c) and time-averaged probability $P_<$  at $t=1000$ (b),(d) {\it versus} modulation frequency $\omega$. Panels (a),(b) correspond to $\gamma =0.8$, panels (c),(d) correspond to $\gamma =1.5$. The input state is the mode $|1-\rangle$. The red arrows indicate the width of the main tunneling suppression peak defined at the $P_<(T)=0.7$ level. In panel (a) the schematic zoom shows the crossing of the Floquet exponents where localization of the state $|1-\rangle$ occurs. Zooms in panel (c) show the crossing points at which full localization of superpositions of the states $|1-\rangle$ and $|2-\rangle$ is possible. In all cases $f=0.143$. To increase visibility of close curves we use different colors. Regions plotted with magenta color correspond to coincidence of quasi-energies as a result of degeneracy of lower energy levels.}
\label{fig:two}
\end{figure}

We start with a general case of \textit{partial} suppression of tunneling for the lower modes. In Figs.~\ref{fig:two}(a),(b) we compare dependencies of the Floquet exponents (panel a) and time-averaged probability $P_<(t)$ to find atom in the left well (panel b) on modulation frequency $\omega$ for moderate SOC strength $\gamma=0.8$. Clearly, peaks in time-averaged probability indicating on suppression of tunneling correspond to the crossing of two Floquet exponents [schematic zooms in panel (a) highlight these crossings]. There is infinite number of peaks occurring at progressively decreasing $\omega$ values. The peaks in $P_<(t)$ dependence on $\omega$ have finite "width" $\delta\omega(t)$ that we measure at $P_<=0.7$ level. Obviously, $\delta\omega(t)$ is determined by the angle between Floquet exponent curves. The width of the peaks become smaller with decrease of the frequency at which peak occurs. Due to its definition $\delta\omega(t)$ decreases to zero at $t\to\infty$. {\color{black}Notice that frequencies corresponding to peaks increase almost linearly with increase of the amplitude of modulation $f$, but the width of peaks is weakly affected by this amplitude.}

While the agreement between four-mode and continuous models is practically exact for small amplitudes of modulation $f$, for moderate $f$ values the mismatch of a few percents is possible. This is a peculiarity of the time-varying Hamiltonian resulting in limitations of the four-mode model, which does not account for the continuous spectrum. Indeed, the amplitude $f=0.143$ used in Fig.~\ref{fig:two} corresponds to variations of the potential depths that are sufficiently large to locally increase the number of bound states in the potential at instants of its maximal deformations (i.e. at $\omega t=\pi/2+\pi n$). In the basis $|i\pm\rangle$ used for derivation of (\ref{eq:matrix_main}) this means that the continuous spectrum also becomes excited that leads to the increase of the frequency of oscillations as compared with four-mode model (\ref{eq:matrix_main}). Nevertheless, four-mode model gives qualitatively correct and quantitatively  accurate predictions in all the cases reported here.

\begin{figure}[h]
\includegraphics[width=\columnwidth]{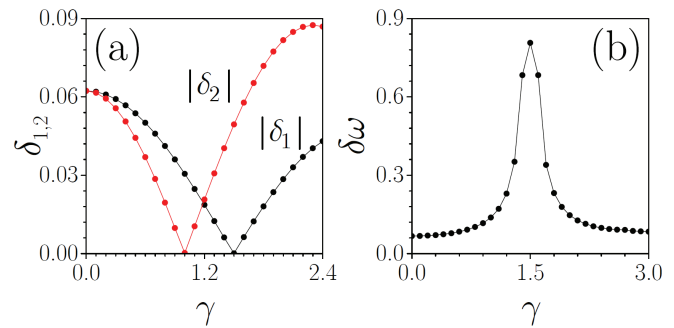}
\caption{(a) Modulus of the energy difference for lower and upper levels versus SOC strength. (b) The full width of the frequency domain in which suppression of tunneling occurs as a function of $\gamma$ at $f=0.143$.}
\label{fig:three}
\end{figure}

One of our central results is that strength of SOC drastically affects suppression of tunneling and it can lead to strong broadening of peaks in the dependence of $P_<(t)$ on $\omega$. Such broadening occurs close to the point, where energy levels in a given pair become degenerate, a SOC-induced phenomenon that was previously encountered in lattices~\cite{SOC-Bloch,KKZT}. In a generic situation for realistic potentials the collapse of lower and upper pairs of energy levels occurs at very different SOC strengths [Fig.~\ref{fig:three}(a)], i.e. in our case lowest levels collapse $(\varepsilon_{11}=\varepsilon_{12})$ at $\gamma\approx 1.5$. In this case $\delta_1=0$ and the states $|1\pm\rangle$ become exact eigenstates of $H_0$. Thus, even in the absence of modulation these lower states remain localized in the respective wells (at the same time the states from upper levels do experience tunneling). This {\em static} suppression of tunneling of lower levels occurs only due to the SOC and is preserved for almost all frequencies in the dynamical regime [see Figs.~\ref{fig:two}(c),(d)]. In Fig.~\ref{fig:two}(c) one observes almost complete overlap of pairs of Floquet exponents, corresponding to plateaus in $P_<(t)$ vs $\omega$ dependence. However, quasi-energies split in the vicinity of the center, $\lambda=0$, and at the boundaries, $\lambda=\pi$, of the "Brillouin zone". Precisely in these points one observes delocalization, manifested in narrow dips in  $P_<(t)$ curve [Fig.~\ref{fig:two}(d)]. These dips limit the width of the frequency domain $\delta\omega$, where effective suppression of tunneling occurs. Figure ~\ref{fig:three}(b) shows the width of this domain (defined for right outermost peak) as a function of SOC strength $\gamma$. Nearly ten-fold broadening of the frequency interval where tunneling is suppressed is obvious. Similar dependence can be obtained for the excitation of only upper levels, but peak will be located at $\gamma=1$.

{\color{black} We emphasize that Fig.~\ref{fig:three} represents a summary of the results obtained for different values of SOC in independent numerical runs (thus applicable to independent experimental settings), rather than for the adiabatic time variations of $\gamma$. Even though variation of $\gamma$ in time may cause heating, in principle, the experiments exist, where this problem does not appear. For estimate one can use $\gamma=k_{\rm SO}/k$, where $\hbar k$ is the linear momentum of the atom, and $k_{\rm SO}$ is the characteristic wavenumber defining SOC strength. In the experiment of Ref.~\cite{SOCtuning} with rubidium atoms, where $k$ can be estimated as $0.2\,\mu$m$^{-1}$ corresponding to a trap with average frequency $2\pi\times 60$ Hz, the quantity  $k_{\rm SO}$ was varying between $0$ and $4.9\,\mu$m$^{-1}$ within 100 ms time interval without noticeable heating. Therefore, SOC range covered in this experiment greatly exceeds the range shown in Fig.~\ref{fig:three}.}

Restoration of tunneling occurs at frequencies [centers of the dips in Fig.~\ref{fig:two}(d)], corresponding to resonant transitions between the lower and upper states localized in the left and right wells, respectively. Such transitions are possible due to spinor character of our system. At such frequencies Josephson-Rabi oscillations occur. In terms of dynamical system (\ref{eq:matrix_main}) this is the case of the parametric resonance occurring at frequencies $\omega=\Delta,\Delta/2,...$ that {\color{black}do not depend on modulation amplitude $f$ (hence, by changing $f$ one can shift frequencies at which tunneling is suppressed, and one can control the width of corresponding frequency intervals with SOC strength $\gamma$, but these intervals remain limited due to resonant transitions)}. Indeed, neglecting $u$ (as explained above) and setting, for example, $\omega=\Delta$, one can look for a solution of (\ref{eq:matrix_main}) in the form  $c_{j\pm}(t)=\exp\left[ (-1)^j\Delta t\right]\tc_{j\pm}(t)$ where $\tc_{j\pm}(t)$ are functions, slowly varying on the period $2\pi/\omega$. Averaging over fast oscillations yields $\tilde{\bc}(t)=\sum_n \alpha_n\tilde{\bc}^{(n)} e^{i\nu_n t/2}$ with four frequencies $\nu_n =\pm[\delta_2\pm(\delta_2^2+f^2w^2)^{1/2}]/2$ ($n=1,...,4$) determining excitations of {\em all four modes}, since  $\alpha_n=(\tilde{\bc}^{(n)})^\dagger\bc(t=0) \neq 0$, where $\tilde{\bc}^{(n)}$ is the orthonormal basis.

Figure~\ref{fig:two}(c) illustrates the possibility of the \textit{full} suppression of tunneling. In a general case with four-level potential and arbitrary SOC strength such supression is impossible, because it would require simultaneous crossing of two pairs of quasi-energies. However, if SOC strength is such that there exists degeneracy of one pair of energy levels (e.g. at $\gamma=1.5$ in our case), full localization can be achieved because one needs to realize crossing of only two Floquet exponents due to temporal modulation. In the case $\gamma=1.5$ this situation is encountered at $\omega\approx 1.165$ in terms of the four-mode model [Fig.~\ref{fig:two}(c)] and at $\omega\approx 1.185$ in the original continuous model. In this case the qbit state $c_{1} |1\pm\rangle+c_2|2\pm \rangle$ remains dynamically localized in the left ("$-$") or right ("$+$") potential wells even though it combines modes from both upper and lower levels.

\begin{figure}[h]
\includegraphics[width=\columnwidth]{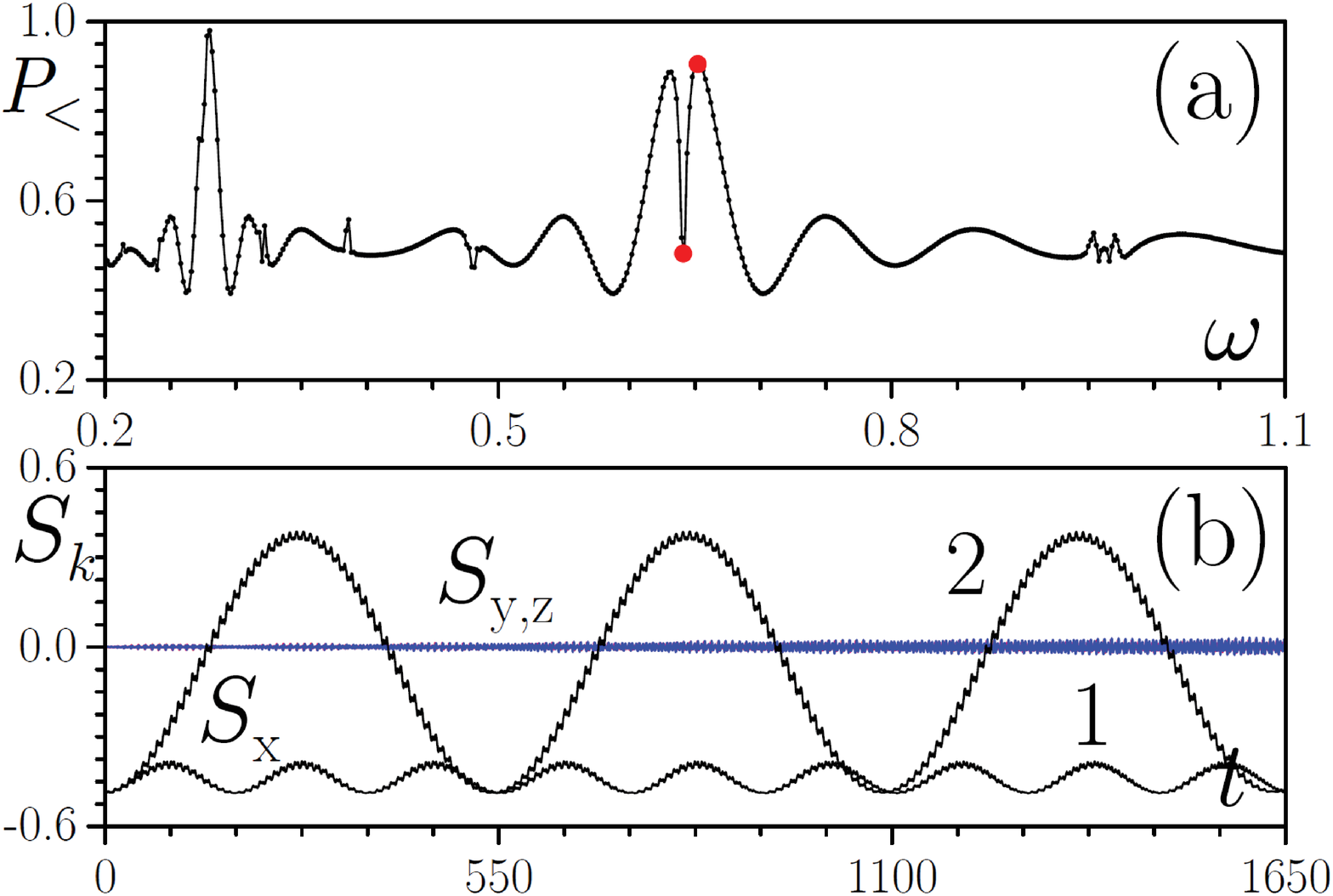}
\caption{(a) Time-averaged probability $P_<$ {\it versus} modulation frequency $\omega$ for $\gamma =0.8$ at $t=1000$. (b) Evolution of $x-$projections of the spin for frequencies marked by the red dots in the localization curve in panel (a), $\omega = 0.652$ (curve 1) and $\omega = 0.640$ (curve 2). $S_{y,z}$ are close to zero and are	shown for one frequency only. The input state is $|1-\rangle$. In all cases $f= 0.0774$.}
\label{fig:four}
\end{figure}

The frequencies of Josephson-Rabi transitions (described by elements $w$ in the four-mode model) are determined by the potential and {\color{black}do not depend on the amplitude $f$ of its modulation}. In contrast, crossing of quasi-energies is controlled by the frequency $\omega$ and amplitude $f$ of modulation. This allows to adjust $f$ to achieve coincidence of the frequency of Josephson-Rabi transitions with frequency of crossing of quasi-energies. When this happens, the respective peak of the localization curve splits into two peaks due to appearance of a narrow delocalization region, where tunneling is enhanced [Fig.~\ref{fig:four}(a)]. In this region tunneling occurs only due to the transitions $|1\pm\rangle \leftrightarrow |2\mp\rangle$, i.e. it is accompanied by the spin flip of the system. Localization of spin in one potential well corresponding to the red dot in the maximum of $P_<(t)$ curve is shown by curve 1 in Fig.~\ref{fig:four}(b), while spin-flipping due to the Josephson-Rabi oscillations corresponding to the red dot in the minimum of $P_<(t)$ is shown by curve 2. We observe that the flipping occurs through the mixture of states induced by periodic modulations: only $S_x$ is changing, while other two mean projections remain nearly zero $|S_{y,z}|\ll 1/2$.

\section{On effect of weak nonlinearity}
\label{sec:nonlinearity}

\begin{figure}[h]
\includegraphics[width=\columnwidth]{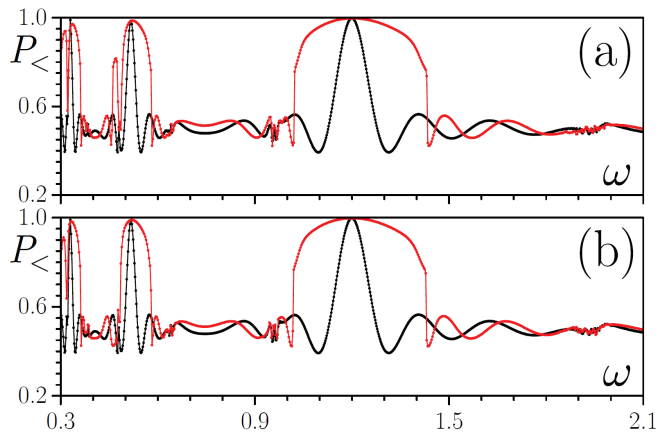}
\caption{{\color{black}(a) Time-averaged number of condensed atoms $P_<$ {\it versus} modulation frequency $\omega$ at $\gamma =0.8$, $f=0.143$, and $t=1000$ in the presence of weak attractive $g=-0.02$ (a) or repulsive $g=+0.02$ (b) interactions (red curves). Black curves in both panels correspond to $g=0$ and are shown to stress broadening of peaks caused by nonlinear interactions. The input state is $|1-\rangle$.}}
\label{fig:five}
\end{figure}

{\color{black} Finally, we addressed the impact of weak inter-atomic interactions on dynamical suppression of tunneling. The respective results can be viewed as a suppression of tunneling of a SO-BEC which solves the Gross-Pitaevskii equation 
\begin{eqnarray}
i\frac{\partial \bPsi}{\partial t}=H\bPsi+g(\bPsi^\dagger \bPsi)\bPsi
\end{eqnarray}
 in which the nonlinearity coefficient $g$ describes inter-atomic interactions. Now $P_<(\omega)$ acquires the meaning of the total number of atoms localized in the domain $x<0$ (i.e., as a matter of fact in the left potential well).  When SOC strength $\gamma$ is far from the value at which collapse of energy levels occurs, the effect of dynamical suppression is strongly sensitive to nonlinear interactions. Already for small values of $|g|\sim 0.02$ both attractive and repulsive interactions lead to noticeable broadening of all peaks in $P_<(\omega)$ dependence, i.e. in this regime the interactions favor suppression of coupling independently of its sign [compare panels (a) and (b) in Fig.~\ref{fig:five}]. For small $|g|$ values main peak broadens nearly by the same factor for both attractive and repulsive interactions. However, already at $g\sim 0.2$ repulsive interactions lead to appearance of broad frequency intervals, where coupling is enhanced, while for attractive nonlinearity $g\sim -0.2$ it is suppressed practically for all modulation frequencies. In contrast, when SOC strength is close to the value at which crossing of energy levels occurs, i.e. $\gamma=1.5$, coupling remains suppressed even for relatively strong repulsive interactions with $g\sim 0.2$. Such nonlinearity, however, substantially affects frequency values at which $P_<(\omega)$ acquires minimal values (i.e. coupling is enhanced).}

\section{Conclusions}

{\color{black}In this paper we studied dynamics of a spin orbit atom in a double well trap with out-of-phase oscillating depths of the potential wells. Some results are also presented for interacting spin-orbit coupled Bose-Einstein condensates. The explored configurations of the potential correspond to symmetric potential wells, each of which supports two eigenstates when it is isolated. We have found different types of evolution. They include partial dynamical suppression of tunneling for a single input state, as well as full suppression characterized by simultaneous localization of two different input states in a single well. Full suppression of tunneling is a result of simultaneous action of two different effects: energy levels crossing in static potential and crossing of the Floquet exponents (quasi-energies) in the modulated potential.

 It was shown that by tuning strength of the spin-orbit coupling one can drastically expand frequency domains where suppression of tunneling is possible. We have compared the four-mode discrete model with the original continuous one, and found good qualitative agreement, however with appreciable (of the order of a few percents) quantitative differences in the resonant frequencies.

The reported system allows to implement controlled localization of a qbit in one potential well, that can be used to manipulate switching of the average spin between potential wells or to design a pseudo-spin splitter for single atoms as well as for spin-orbit coupled Bose-Einstein condensates. Suppression of tunneling for SOC atoms in periodic potentials and its interplay with Landau-Zener tunneling is also among potential extensions of the results reported here.}
  
\acknowledgments
Y.V.K. acknowledges support from the Severo Ochoa program (SEV-2015-0522) of the Government of Spain, Fundacio Cellex, Generalitat de Catalunya and CERCA, as well as partial support by the program 1.4 of Presidium of RAS "Topical problems of low temperature physics". V.V.K. was supported by the FCT (Portugal) under Grant No. UID/FIS/00618/2013.

\end{document}